\PassOptionsToPackage{svgnames}{xcolor}
\documentclass[a4paper,10p,runningheads]{llncs}
\usepackage[english]{babel}

\usepackage{xurl}
\usepackage{hyperref}
\usepackage[T1]{fontenc}
\usepackage{times}
\usepackage[scaled=0.81]{beramono}

\makeatletter
\newcommand\niceparagraph{\@startsection{paragraph}{4}{\z@}{-4\p@ \@plus -4\p@ \@minus -4\p@}{-0.5em \@plus -0.22em \@minus -0.1em}{\normalfont\normalsize\bfseries}}
\newcommand\nicesubsubsection{\@startsection{subsubsection}{3}{\z@}{6\p@ \@plus -0.5\p@ \@minus -0.5\p@}{-0.5em \@plus -0.22em \@minus -0.1em}{\normalfont\normalsize\bfseries\boldmath}}
\makeatother
\renewcommand{\subsubsection}[1]{\nicesubsubsection{#1.}}
\renewcommand{\paragraph}[1]{\niceparagraph{#1.}}

\usepackage{cite}

\usepackage{savesym}
\savesymbol{endnote}

\usepackage{booktabs}
\usepackage{listings}
\usepackage{lstautogobble}

\lstdefinestyle{displayed}{
  numbers=none, firstnumber=last, 
  breaklines=true,
  breakatwhitespace=false,         
  keepspaces=true,                 
  numbersep=2pt,
  showspaces=false,                
  showstringspaces=false,
  showtabs=false,                  
frame=tb,
  captionpos=b,                    
numberstyle=\scriptsize,
  tabsize=2,
  captionpos=b,
xleftmargin=0mm, xrightmargin=0mm, basicstyle=\ttfamily\small,
  keepspaces=true,
  columns=fixed,
  escapeinside={(*}{*)},
  mathescape=true,
  showstringspaces=false,
keywordstyle=\color{blue}\bfseries\ttfamily,
  commentstyle=\color{DarkGreen}\itshape\ttfamily,
  ndkeywordstyle=\color{purple}\bfseries,
  identifierstyle=\color{black},
  stringstyle=\color{darkgray}\ttfamily,
}

\lstdefinestyle{displayedC}{style=displayed,escapeinside={($}{$)}}
\lstdefinestyle{displayedPy}{style=displayed,mathescape=false}

\lstdefinelanguage{JavaScript}{
  morekeywords=[1]{break, continue, delete, else, for, function, if, in,
    new, return, this, typeof, var, void, while, with},
morekeywords=[2]{false, null, true, boolean, number, undefined,
    Array, Boolean, Date, Math, Number, String, Object},
morekeywords=[3]{eval, parseInt, parseFloat, escape, unescape},
  sensitive,
  morecomment=[s]{/*}{*/},
  morecomment=[l]//,
  morecomment=[s]{/**}{*/}, morestring=[b]',
  morestring=[b]"
}[keywords, comments, strings]

\newcommand{\Jc}[1]{\mbox{\lstinline[basicstyle=\ttfamily\footnotesize,language=Java]|#1|}}
\newcommand{\JSc}[1]{\mbox{\lstinline[basicstyle=\ttfamily\footnotesize,language=JavaScript]|#1|}}
\newcommand{\Sc}[1]{\mbox{\lstinline[basicstyle=\ttfamily\footnotesize,language=Scala]|#1|}}
\newcommand{\Cc}[1]{\mbox{\lstinline[basicstyle=\ttfamily\footnotesize,language=C]|#1|}}
\newcommand{\Pc}[1]{\mbox{\lstinline[basicstyle=\ttfamily\footnotesize,language=Python,mathescape=false]|#1|}}

\usepackage{enotez}

\restoresymbol{enotez}{endnote}
\setenotez{list-name={URL References}}
\setenotez{counter-format=alph}
\setenotez{mark-cs={\formatEndNoteMark}}
\setenotez{list-heading=\section*{#1}}
\newcommand{\formatEndNoteMark}[1]{\textsuperscript{\{\color{blue}{\textsf{#1}}\}}}

\DeclareDocumentCommand{\URLref}{O{} m}
{\enotezendnote{\IfNoValueF{#1}{#1\xspace}\url{#2}}}

\usepackage{graphicx}

\usepackage{dsfont}
\usepackage{latexsym}
\usepackage{csquotes}
\usepackage{tcolorbox}
\usepackage{listings}
\usepackage{float}
\usepackage{multirow}
\usepackage[scaled]{helvet}
\usepackage[noend]{algpseudocode}
\usepackage{mathrsfs}
\usepackage[linesnumbered, ruled, lined]{algorithm2e}  \usepackage{mathpartir}
\usepackage{mathtools}
\usepackage{stmaryrd}
\usepackage{textcomp} 
\usepackage{bbm}
\usepackage{verbdef}
\usepackage{xspace}
\usepackage{verbatim}
\usepackage{lipsum}
\usepackage{wrapfig}

\usepackage[shortlabels,inline]{enumitem}
\usepackage{pifont}
\usepackage{varwidth}
\usepackage{xpatch}
\usepackage{multirow}
\usepackage{xcolor}
\usepackage[export]{adjustbox}
\usepackage{caption}
\usepackage{subcaption}
\usepackage{setspace}
\usepackage[capitalize]{cleveref}
\usepackage{makecell}
\usepackage{dashrule}
\usepackage{comment}
\usepackage{arydshln}
\usepackage{ifsym}
\usepackage{mdframed}
\usepackage{fancyvrb}

\setlist[enumerate]{label=\emph{\roman*})}
\setlist[description]{font=\normalfont\bfseries}

\usepackage{flushend}

\usepackage{tikz}
\usetikzlibrary{calc,positioning,math,arrows,shapes,fit,intersections}
\usepackage{pgfplots}
\pgfplotsset{compat=newest}

\usepackage{xstring,xparse}

\usepackage{fontawesome}

\newif\ifdraft
\draftfalse
\ifdraft
\usepackage[colorinlistoftodos,textsize=scriptsize,obeyFinal,textwidth=33mm]{todonotes}

\DeclareDocumentCommand{\ReviewNote}{s o m O{white}}{\todo[color=#4,\IfBooleanTF{#1}{inline}{}]{\IfNoValueF{#2}{\textbf{#2:}\xspace}#3}
}
\else
\DeclareDocumentCommand{\ReviewNote}{s o m O{white}}{}
\fi

\definecolor{acol}{HTML}{d7191c}
\definecolor{bcol}{HTML}{fdae61}
\definecolor{ccol}{HTML}{ffffbf}
\definecolor{dcol}{HTML}{abd9e9}
\definecolor{ecol}{HTML}{2c7bb6}

\definecolor{fcol}{HTML}{a6611a}
\definecolor{gcol}{HTML}{dfc27d}
\definecolor{hcol}{HTML}{80cdc1}

\title{Challenges of\\ Multilingual Program Analysis}

\title{Multilingual Program Analysis:\\Challenges and Opportunities}

\title{Challenges of Multilingual\\ Program Analysis and Specification}

\title{Challenges of Multilingual\\ Program Specification and Analysis\thanks{Work partially supported by SNF grant 200021-207919 (LastMile).}}

\author{Carlo A.\ Furia \and Abhishek Tiwari}
\institute{
  Software Institute, USI Università della Svizzera italiana, Switzerland\\
  \email{\url{bugcounting.net}$\quad\cdot\quad$\url{abhishek.tiwari@usi.ch}}
}

\begin{document}

\renewcommand{\sectionautorefname}{Section}
\renewcommand{\subsectionautorefname}{Section}
\renewcommand{\figureautorefname}{Figure}
\renewcommand{\tableautorefname}{Table}

\maketitle

\begin{abstract}
  Multilingual programs,
  whose implementations are made of different languages,
  are gaining traction especially in domains,
  such as web programming, that particularly benefit
  from the additional flexibility brought by using multiple languages.
  In this paper, we discuss the impact that the features commonly used in multilingual programming have
  on our capability of specifying and analyzing them.
  To this end, we first outline a few broad categories of multilingual
  programming, according to the mechanisms that are used for
  inter-language communication.
  Based on these categories, we describe several instances
  of multilingual programs, as well as the intricacies
  that formally reasoning about their behavior would entail.
  We also summarize the state of the art
  in multilingual program analysis, including the
  challenges that remain open.
  These contributions can 
  help understand the lay of the land in
  multilingual program specification and analysis,
  and motivate further work in this area.
\end{abstract}

\section{Introduction}
\label{sec:introduction}

It is increasingly common that modern programming frameworks
offer \emph{multilingual} features,\footnote{Multilingual programs are also sometimes called \emph{polyglots}.}
where different modules of an application are written
using different programming languages.
The rise of web development,
where an application's front-end and back-end are often developed
using different languages,
has certainly made multilingual programming commonplace;
however, various forms of multilingual programs
have been around for much longer (for example,
most high-level programming languages offer some form
of foreign-function interface).

Regardless of the details of how it is implemented,
multilingual programming is likely to add to the challenges
of rigorously specifying and analyzing a program's behavior.
The programming languages that cooperate in a multilingual program
often have (markedly or subtly) different features
such as types, memory models, paradigms, and idioms.
Even in the optimistic scenario where
detailed formal specifications of each component are available,
combining them at the language boundary
to analyze a multilingual program's overall behavior
requires additional technical insights.

This paper aims at drawing a high-level picture
of how program specification and analysis
are complicated by multilingual features,
in order to both better understand the main open challenges,
and to motivate further research in this area.
To this end, it
first discusses, in \autoref{sec:dimensions},
various mechanisms
for inter-language communication that are commonly used
in different kinds of multilingual programs.
For each of them, \autoref{sec:examples} outlines
the main challenges that they pose
to (formally) specifying and analyzing programs using them,
and demonstrates them, in a nutshell, on simple examples.
The examples are stripped down to be as clear and as succinct as possible;
however, they recapitulate challenges and issues
that are present, on a much larger scale,
in real-world programs.
\autoref{sec:soa} and \autoref{sec:discussion} bear out this intuition
by discussing the essential state of the art
of multilingual program analysis,
and outlining its open challenges.

\section{Dimensions of Multilingual Programming}
\label{sec:dimensions}

We organize our discussion of multilingual program analysis
along two dimensions.
First, \autoref{sec:levels} presents a broad categorization of
multilingual \emph{programs} according to the \emph{level} of abstraction
that their inter-language communication targets.
Then, \autoref{sec:layers} introduces
\emph{layers} of program \emph{specification}
(and the corresponding \emph{analyses} that they enable)
that are relevant for multilingual programs.

As for the rest of this paper,
the dimensions are not meant to be exhaustive or strictly mutually exclusive.
Pragmatically,
they are useful to illustrate the most common challenges of
multilingual program specification
on concrete, relevant examples.

\subsection{Multilingual Program Levels}
\label{sec:levels}

One natural way to classify multilingual programs
is according to the features multilingual communication relies on.
We identify four groups of features,
going roughly from higher-level to lower-level.
This classification takes the perspective of a programmer
developing an application in a host language that includes parts written in a guest language; thus, the ``higher-level'' the multilingual features are
the more transparent and ``natural''
the combination of functionalities written in the two languages is.

\begin{description}
\item[API level.] When the host language offers an \emph{API}
  with features to directly access the guest language's runtime,
  writing host code that interacts with the guest is straightforward,
  as it is very similar to using a regular library in the host language.
  For example, as we show in \autoref{sec:ex:API},
  Android apps (implemented in Java or Kotlin)
  may define JavaScript interfaces that give direct access to the app's functionality.

\item[IR level.] If the host and the guest languages are deployed on the same
  runtime, it means their compilers can translate language features
  into a common intermediate representation (\emph{IR}). This provides
  an indirect way of communicating between the two source languages,
  so that a language's code
  can usually call into code written in the other language.
  For example, JVM languages such as Java, Kotlin, and Scala
  all translate into the same bytecode instruction set;
  hence, a program written in, say, Scala can use compiled Java libraries
  with hardly any restrictions---as we discuss in \autoref{sec:ex:IR}'s example.
  
\item[Native level.] A \emph{native} interface gives the host language access
  to native applications---that is, applications that run natively on the host's hardware and operating system.
  For example, as we show in \autoref{sec:ex:native},
  JNI (Java Native Interface) offers direct access
  to C or even assembly code from Java by means of suitable conversion
  functions and wrappers.
  While native multilingual programs can be considered a special case
  of those relying on an API,
  the expressiveness gap between a high-level language (such as Java)
  and a system-specific binary is likely to be wider than
  the gap between high-level languages plugged into a shared API.

\item[System level.] The last group of features for multilingual programming
  is very broad, as it includes all cases of communication using
  lower-level communication primitives offered by the \emph{system}
  that is running the program.
  For example, two programs running on the same operating system
  can communicate
  through inter-process communication;
  even if they run on two different machines,
  they can still communicate by means of networking primitives.
\end{description}

\subsection{Analysis and Specification Layers}
\label{sec:layers}

Consider the analysis of a multilingual program
written in a combination of a host language
and a guest language.
Conceptually,
the interface between host and guest
is summarized by a (implicit or explicit, inferred or expressed) \emph{specification}.
A certain kind of program \emph{analysis}
can cross the inter-language interface
only if
the interface specification
captures the right kind of information.
Accordingly, we consider three levels of specification layers---roughly going from less to more detail.

\begin{description}
\item[Types.] The \emph{types}
  of a procedure's input and output
  are a basic constrain on its valid behavior.
  Thus, they are a natural way of exchanging information across
  language boundaries---as well as a possible source of subtle misbehavior
  when converting between similar types of two languages
  may lose information.

\item[Dataflow.]
  If a dataflow analysis can cross the interface between two languages,
  it means that it can follow program paths that go from one language and the other.
  It is well known that dataflow analysis is especially convenient to verify
  a range of \emph{security} properties such as noninterference,
  as well as to detect common programming errors such as resource \emph{leaks}.

\item[Effects.]
  Fully specifying the \emph{side effects} of a call
  (in other words, its \emph{frame})
  is a challenge even for individual
  programming languages~\cite{SummersDrossopoulouMueller09a,DBLP:conf/ecoop/LeinoM04,PTFM-FM14-SemiCola,DBLP:conf/birthday/BoerG15}.
  The ``mismatch'' that often occurs between
  the semantics of different languages
  renders fully expressing, and reasoning about,
  the effects of an inter-language call
  especially daunting.
\end{description}

\section{Examples of Multilingual Analysis and Specification}
\label{sec:examples}

This section presents several examples of snippets of code
from multilingual programs, and discusses the ensuing
challenges of analysis and specification on them.

\subsection{API Level}
\label{sec:ex:API}

\begin{figure*}[!tb]
  \centering
\begin{subfigure}{0.47\textwidth}
\begin{lstlisting}[firstline=16,lastline=29,language=Java,numbers=left]
class Messenger {
  private String message;

  @JavascriptInterface
  public String getMessage();

  @JavascriptInterface 
  public String getInfo();
  @JavascriptInterface
  public void setInfo(String info);

  @JavascriptInterface
  public ArrayList<String> asList();
}
\end{lstlisting}
    \caption{An example of a Java class whose interface methods are
      made available to JavaScript code
      through Android's WebView component.}
    \label{fig:ex:api:android-interface}
  \end{subfigure}
\begin{subfigure}{0.52\textwidth}
\begin{lstlisting}[firstline=1,lastline=14,language=Java,numbers=right]
class MainActivity {
  protected void onCreate(Bundle state)
  {
    WebView wv = new WebView(this);(*\label{l:wv:createwv}*)
    WebSettings wvset = wv.getSettings();
    // enable JavaScript bridge
    wvset.setJavaScriptEnabled(true);(*\label{l:wv:enableJS}*)
    // add interface object
    Messenger msg = new Messenger(this);(*\label{l:wv:obj}*)
    wv.addJavasriptInterface(msg, "msg");(*\label{l:wv:export}*)
    // run JavaScript from URL or string
    wv.loadUrl(/* URL or JS */);(*\label{l:wv:loadURL}*)
  }
}
\end{lstlisting}
    \caption{An example of Android activity that instantiates a WebView component,
    enables JavaScript communication, and makes an instance of \Jc{Messenger} available to the JavaScript runtime.}
    \label{fig:ex:api:android-activity}
  \end{subfigure}
\begin{subfigure}{0.51\textwidth}
\begin{lstlisting}[firstline=1,lastline=10,language=JavaScript,numbers=left]
// Call function after DOM is loaded
document.addEventListener(
  "DOMContentLoaded", function() {
    var message = msg.getMessage();
    // Add message obtained from Java
    // as a new paragraph in webpage
    var p = document.createElement("p");
    p.textContent = message;
    document.body.appendChild(p);
});
\end{lstlisting}
    \caption{An example of JavaScript code that reads a message from Android/Java and displays it into the current web page.}
    \label{fig:ex:api:js-code}
  \end{subfigure}
\begin{subfigure}{0.48\textwidth}
\begin{lstlisting}[firstline=13,lastline=22,language=JavaScript,numbers=right]
// Information security leak
var device = msg.getInfo();(*\label{l:js:read}*)

// Dataflow from JS to Android
msg.setInfo((*\label{l:js:write}*)
  document.getElementById("Search")
);

var lst = msg.asList();(*\label{l:js:object}*)
// lst is a generic object(*\label{l:js:object-2}*)
\end{lstlisting}
    \caption{Snippets of JavaScript code that misuse the Java-JavaScript bridge by modifying an Android app's behavior in unintended ways.}
    \label{fig:ex:api:js-snippets}
  \end{subfigure}
\caption{Code snippets that demonstrate Java-JavaScript hybrid programs
  written using Android's WebView framework.}
  \label{fig:ex:api}
\end{figure*}

As a significant example of multilingual programming through an API,
we consider \emph{hybrid} mobile applications (``apps'')
that use
Android's WebView
component.\footnote{
  In the paper, we refer to several online documents
by means of URL references:
these are marked by superscript letters in blue
between curly braces,\URLref[An example of URL reference:]{https://bugcounting.net}
so that they can be easily distinguished from regular footnotes,
and listed at the end of the paper after the usual bibliographic references.
}\URLref{https://developer.android.com/reference/android/webkit/WebView}
This component
embeds a headless browser within a host Android app;
the browser can, among other things, run JavaScript code
that accesses the Android runtime through a programmer-defined API.

\autoref{fig:ex:api:android-interface}
shows a Java class \Jc{Messenger} that marks four methods
with the annotation \Jc{@JavascriptInterface}.\URLref{https://developer.android.com/reference/android/webkit/JavascriptInterface}
In \autoref{fig:ex:api:android-activity},
the app's main activity
initializes the WebView component (line~\ref{l:wv:createwv}),
allows it to run JavaScript code (line~\ref{l:wv:enableJS}),
creates an instance \Jc{msg} of the \Jc{Messenger} class (line~\ref{l:wv:obj}),
and registers it with the WebView object:
after line~\ref{l:wv:export},
any JavaScript code that runs in the WebView will
have access to a persistent property \JSc{msg}
of the browser's DOM instance with an interface
corresponding to class \Jc{Messenger}'s \Jc{@JavascriptInterface} methods.
Finally, in line~\ref{l:wv:loadURL}, 
the Android activity loads content in the WebView
by passing either the URL of a webpage
(which may embed JavaScript code),
or directly JavaScript code as a string
to \Jc{loadUrl}.
\autoref{fig:ex:api:js-code} is an example of
JavaScript code that could be passed, as a string,
to \Jc{loadUrl}: a callback
that, as soon as the DOM is fully loaded,
calls \JSc{getMessage()} on the bridged object \JSc{msg}
and inserts the received string as a new paragraph of the current
web page.

The Java-JavaScript bridge offered by Android's WebView component
is widely used in mobile apps~\cite{LuDroid-Journal, mutchler2015large}\URLref[Mobile Developer Survey]{https://ionicframework.com/survey/2017\#trends},
as it provides a convenient way of
programming plat\-form-independent web components in JavaScript,
and then combining them with Java code to create a full-fledged app.
Using the WebView's API-level inter-language communication framework,
developers can flexibly define the Java interface that will be
visible and accessible to the JavaScript code (\autoref{fig:ex:api:android-interface}),
co-designing the components written in the two languages
so that they seamlessly communicate.
On the other hand, this flexibility may also complicate
the analysis and specification of the inter-language program behavior,
as we'll discuss henceforth.

\paragraph{Types}
Any Java method can be annotated with \Jc{@JavascriptInterface}.
However, Java's and JavaScript's type systems are quite different:
only Java's primitive and string types are reliably converted to
their JavaScript counterparts
(e.g., numeric, Boolean, and string types). Consequently, multiple studies~\cite{hybridDroid, hydridDroidICSE2019} have shown several instances of type bugs in Java-Javascript communication in hybrid apps. Besides, instances of complex (possibly user-defined) Java reference types
appear in JavaScript as generic \JSc{object} instances,
as outlined in \autoref{fig:ex:api:js-snippets}'s
lines~\ref{l:js:object}--\ref{l:js:object-2}.
Thus, even if the Java-JavaScript communication
goes through a user-defined interface with typed signatures,
fully analyzing the overall behavior of a hybrid app
requires a precise formalization of type conversions
between two fundamentally different type systems---in particular, static vs.~dynamic.
In fact, some JavaScript operations behave slightly differently
depending on whether they run on ``native'' JavaScript objects
or on Java-bridged objects;
for example, Tiwari et al.~\cite{TiwariPH23}
noticed that deleting an instance method of a Java interface object
has no effect---whereas one can delete individual methods of
a regular JavaScript object.

\paragraph{Dataflow}
The information flow in Android hybrid apps is
\emph{bidirectional}:
the JavaScript runtime can not only read
information from the Java runtime
(line~\ref{l:js:read} in \autoref{fig:ex:api:js-snippets})
but also write it back
(line~\ref{l:js:write} in \autoref{fig:ex:api:js-snippets}).
This compounds the difficulty of rigorously analyzing the dataflow
that cross language boundaries
(which is already challenged~\cite{7194584, pradel2015good,  10.1145/3106741} by JavaScript's highly dynamic nature\URLref{https://developer.android.com/develop/ui/views/layout/webapps/webview\#BindingJavaScript}).
For example,
many Android techniques to detect information-flow security leaks~\cite{hybridDroid, TiwariPH23}
are based on the idea of detecting flows of sensitive information
into public channels.
There is evidence~\cite{LuDroid-Journal, 10172950}
that several Android hybrid apps
suffer from such information-flow security issues\URLref{ https://github.com/gustavogenovese/androidSamples/blob/036c1c6f9bc35392d2c18d6876073b13c31a47b8/webview/src/main/java/com/gustavogenovese/webview/SynchronousJavascriptInterface.java#L41}
that state-of-the-art
analysis techniques would not be capable of detecting automatically.

\paragraph{Effects}
Line~\ref{l:js:write} in \autoref{fig:ex:api:js-snippets}
also captures a scenario where detecting all the effects
of executing some JavaScript code within a WebView component
would be challenging.
The JavaScript code sets some ``information'' state component in the
Android runtime to the search results in the current web page;
since the search results are loaded dynamically,
generally specifying how the state will change
requires some kind of environment specification.
Studies~\cite{LuDroid-Journal, 10172950} have found several examples of apps\URLref{https://apkcombo.com/ko/mas-que-panes-y-peces/com.a2stacks.apps.app57191abb7ab09/}\URLref{https://apkcombo.com/zh/sally-s-makeup-salon/com.nuttyapps.sally.makeup.salon/}\URLref{https://www.9apps.com/ar/android-apps/Social-network-circus/}
that abuse this mechanism to inject
malware into an Android app,
or to simply obfuscate its behavior.
Android WebView's threaded model
further complicates the synchronization
between the Java and JavaScript runtimes:
JavaScript code may execute asynchronously
with respect to when it is started by an Android call
to \Jc{loadURL},
and hence it may access a bridged object in different states in different executions~\cite{TiwariPH23}.
By itself,
JavaScript's highly dynamic nature
also complicates reasoning about its semantics statically;
for example,
the JavaScript code running in a WebView component
can redefine a method of the JavaScript interface~\cite{TiwariPH23}.

\subsection{IR Level}
\label{sec:ex:IR}

\begin{figure*}[!tb]
  \centering
\begin{subfigure}{\textwidth}
\begin{lstlisting}[firstline=2,lastline=8,language=Scala]
class JListWrapper[A](val underlying: java.util.List[A])
  extends mutable.AbstractBuffer[A]
  with ...  
{
  def length = underlying.size
  def insert(idx: Int, elem: A): Unit = underlying.subList(0, idx).add(elem)
  // ...
\end{lstlisting}
    \caption{An excerpt of Scala's \Sc{scala.collection.convert.JListWrapper}.}
    \label{fig:ex:ir:scala-wrapper}
  \end{subfigure}
\begin{subfigure}{\textwidth}
\begin{lstlisting}[firstline=30,lastline=34,firstnumber=1,numbers=left,language=Scala]
  // Java list, unmodifiable
  val jl: java.util.List[String] = java.util.List.of("a", "b", "c")(*\label{l:ir:listof}*)
  // Convert Java List to Scala mutable Buffer using JListWrapper
  val sl: mutable.Buffer[String] = jl.asScala(*\label{l:ir:asScala}*)
  sl.insert(1, "X")(*\label{l:ir:unsupported}*)
\end{lstlisting}
    \caption{A mismatch between Scala and Java types causes an \Sc{UnsupportedOperationException}.}
    \label{fig:ex:ir:breaking-contract}
  \end{subfigure}
\caption{Code snippets that demonstrate Scala data structures wrapping Java ones.}
  \label{fig:ex:ir}
\end{figure*}

Scala is a modern, statically typed programming
language~\cite{Scala-book}
that runs of the Java Virtual Machine (JVM)
and compiles to bytecode (JVM's intermediate representation).
Scala has prominent \emph{functional} features,
but is actually a multi-paradigm language---also
offering a sophisticated support for object orientation.
Interoperability with Java is another key feature of Scala,
whose programs can seamlessly use
libraries written in other JVM languages.

As a significant example of multilingual Java-Scala programs,
we consider how Scala collections
support Java's widely used \Jc{java.util.List} type.\footnote{
  Scala has substantially revised its collections framework
  with version 2.13 of the language;
  in this paper, we consider Scala~3, which also uses v.~$\geq$2.13
  of the collections.
}
To optimize performance,
when converting a Java collection
to a suitable Scala type,
Scala \emph{wraps} the Java objects with a suitable Scala interface---avoiding, as much as possible, copying the collection's content.\URLref{https://docs.scala-lang.org/overviews/collections-2.13/conversions-between-java-and-scala-collections.html}
\autoref{fig:ex:ir} shows a small excerpt of Scala's wrapper
for Java's \Jc{List},\URLref{https://github.com/scala/scala/blob/06a7509e3a9083793038bf4449281491a8614eb0/src/library/scala/collection/convert/JavaCollectionWrappers.scala\#L138}
as well as an example of Scala code seamlessly using Java lists.
We will now go into the details of this code, and its implications
for the correctness of multilingual programs.

\paragraph{Types}
Scala collections have been carefully designed to ease interoperability with Java
while offering more expressive, functional features on top.
However, this also entails that some limitations of Java collections' design
leak into Scala---rendering a precise conversion between the two type systems
not always possible.
As shown in \autoref{fig:ex:ir:scala-wrapper},
Java \Jc{List} type's wrapper \Sc{JListWrapper}
extends Scala's
\Sc{mutable.AbstractBuffer}.\URLref{https://www.scala-lang.org/api/2.12.6/scala/collection/mutable/AbstractBuffer.html}
This makes perfect sense, since a Java's list interface
provides methods
``to efficiently insert and remove [...] elements
at an arbitrary point in the list''.\URLref{https://docs.oracle.com/en/java/javase/17/docs/api/java.base/java/util/List.html}
The catch is that
Java uses the same \Jc{List} type also to denote \emph{unmodifiable} lists;
calling modification operations on an unmodifiable list is allowed by the type system
but results in an exception at runtime.
This shortcoming of the Java collection API can also affect, through the wrappers,
Scala collections.

The Scala code in \autoref{fig:ex:ir:breaking-contract}'s line~\ref{l:ir:listof}
creates an instance \Jc{jl} of \Jc{java.util.List}
by calling method
\Jc{List.of},\URLref{https://docs.oracle.com/en/java/javase/11/docs/api/java.base/java/util/List.html\#of(E...)}
which returns an unmodifiable list.
Since \Jc{jl}'s immutability is unknown statically,
line~\ref{l:ir:asScala} creates a \emph{mutable} wrapper \Sc{sl} around \Sc{jl}.
Henceforth, calling any modification operation---such as \Sc{insert} at line~\ref{l:ir:unsupported},
which calls \Jc{List.add} on the underlying \Jc{jl}---results in an \Sc{UnsupportedOperationException} at runtime.
This example shows that, even for closely interoperable languages such
as Scala and Java,
precisely reasoning about type conversions at the language interface
may be arduous.

\paragraph{Dataflow}
Dataflow analyses are naturally defined
at the level of an intermediate representation~\cite{appel,dragon-book}.
Thus, combining Java and Scala dataflow analyses at the level of bytecode
is, in principle, an effective way of applying such kind of analysis
to a multilingual program.
Still, designing a static analysis that is effective and sufficiently precise
for programs spanning two different programming paradigms remains challenging.
For example, the popular IntelliJ IDEA includes a dataflow analysis\URLref{https://blog.jetbrains.com/idea/2018/01/fumigating-the-idea-ultimate-code-using-dataflow-analysis/}\URLref{https://www.jetbrains.com/help/idea/analyzing-data-flow.html\#analyze-stack-traces}
that can uncover common programming mistakes such as dead code and null dereferencing.
This functionality was originally developed for Java as a purely intraprocedural analysis;
more recently, it has been extended to Scala code.\URLref{https://blog.jetbrains.com/scala/2021/10/28/dataflow-analysis-for-scala/}
The Scala dataflow analysis is, however, \emph{inter}procedural,
because the IntelliJ developers
decided that being able to follow a control flow into function calls
is paramount in a functional language like Scala;\URLref{https://blog.jetbrains.com/scala/2021/10/28/dataflow-analysis-for-scala/\#interprocedural-analysis}
it is also, arguably, more feasible, by taking advantage of the prevalent usage
of immutable data in Scala---as opposed to old-fashioned imperative programming in Java.

\paragraph{Effects}
The imperative/functional abstraction gap also affects
how we can reason about the precise behavior of a call
where callee and caller are programmed in two different languages.
A strict functional Scala program
consists of a collection of \emph{pure} (side-effect free) functions,
which simplifies reasoning about their behavior.
Scala's expressive type system can still accommodate meaningful side effects
using, for example, an effect system~\cite{DBLP:conf/ecoop/RytzAO13,DBLP:conf/ecoop/RytzOH12}.
However, Scala programs often take full advantage of the language's versatility
by combining purely functional and imperative features---for instance,
to implement parallel and distributed programming features~\cite{DBLP:conf/icse/PankratiusSG12,DBLP:journals/corr/HallerA17}.
As a result, fully analyzing the behavior
of a combination of imperative and functional features
remains a formidable challenge~\cite{DBLP:conf/tools/NordioCMMT10,DBLP:journals/pacmpl/WolffBMMS21}.

\subsection{Native Level}
\label{sec:ex:native}

\begin{figure*}[!tb]
  \centering
\begin{subfigure}{\textwidth}
\begin{lstlisting}[firstline=1,lastline=15,language=Java,numbers=left]
class Messenger {
 private String message;

 public static native String fromC();
 public static native void toC(String msg);

 // load shared library within static block
 static { System.loadLibrary("Messenger"); }

 public String getMessage()
 { message = fromC(); return message; }

 public void setMessage(String msg)
 { message = msg; toC(message); }
}
\end{lstlisting}
    \caption{An example of a Java class some of whose interface methods
      are implemented as native code.}
    \label{fig:ex:native:java-interface}
  \end{subfigure}
\begin{subfigure}{\textwidth}
\begin{lstlisting}[firstline=1,lastline=14,language=C,,style=displayedC,numbers=left]
JNIEXPORT void JNICALL toC (JNIEnv * env, jclass obj, jstring jstr) {
{ const char *cstr;
  cstr = (*env)->GetStringUTFChars(env, jstr, NULL);
  if (cstr == NULL) return;

  // Use cstr as a null-terminated C string ...
  // ...

  // Release memory after usage
  (*env)->ReleaseStringUTFChars(env, str, cstr); }

JNIEXPORT jstring JNICALL fromC (JNIEnv * env, jclass obj)
{ char cstr[] = /* Get string from C program */;
  return (*env)->NewStringUTF(env, cstr); }
\end{lstlisting}
    \caption{Outline of the C implementations JNI native methods \Cc{toC} and \Cc{fromC}.}
    \label{fig:ex:native:c-implementation}
  \end{subfigure}
\caption{Code snippets that demonstrate Java Native Interface's (JNI) capabilities.}
  \label{fig:ex:native}
\end{figure*}

Java Native Interface (JNI) is a framework
that gives Java applications access to native libraries.
JNI has been extensively used
to implement core JDK functionality~\cite{jni-2,jni-1}
such as input/output and device access.
In this section, we go over a simple example of Java-C
hybrid program using JNI, highlighting the features that affect
reasoning about and specifying the multilingual program's behavior.

As mentioned in \autoref{sec:levels},
a multilingual native interface such as JNI
essentially implements a form of API-level communication---but one
with a wide inter-language abstraction gap.
\autoref{fig:ex:native:java-interface} shows a Java class
\Jc{Messenger}
that includes two method signatures \Jc{fromC} and \Jc{toC}
with modifier \Jc{native}.
This denotes that their implementation
is not given in Java, but is to be found in some
native library that is available to the JVM where this code will run.
Apart from this detail,
methods \Jc{fromC} and \Jc{toC}
can be used by the Java program just like any other method---as demonstrated by the other two methods \Jc{getMessage} and \Jc{setMessage}
of the same class \Jc{Messenger}.

A fundamental difference between
\autoref{fig:ex:native}'s
example of Java-C communication,
and \autoref{fig:ex:api}'s example of Java-JavaScript communication
lies in how the two languages communicate at the interface.
In the Java-JavaScript bridge,
methods marked with \Jc{@JavascriptInterface}
are implemented in Java and \emph{callable} from JavaScript;
in contrast,
in the Java-native JNI framework,
methods marked with \Jc{native} in Java are
\emph{implemented} in C (and compiled to native)
and then used regularly in Java.

\paragraph{Types}
\autoref{fig:ex:native:c-implementation}
outlines C implementations
of \autoref{fig:ex:native:java-interface}'s
native methods \Jc{fromC} and \Jc{toC}.
Even in such a simple example,
the wide gap between Java and C types is apparent.
In particular, the C code needs boilerplate code
to explicitly:\URLref{https://github.com/mkowsiak/jnicookbook/}
\begin{enumerate*}
\item access the Java runtime through reference \Cc{env},
\item retrieve the string object pointed to by \Cc{jstr},
\item convert it to a null-terminated C string
  (using JNI utility function \Cc{GetStringUTFChars}).
\end{enumerate*}
Conversely, JNI function \Cc{NewStringUTF} handles
the opposite conversion---from a C character array to a Java string.
On the one hand, these JNI library functions
handle the conversion between corresponding Java and C types,
ensuring that no information is lost.\URLref{https://docs.oracle.com/javase/8/docs/technotes/guides/jni/spec/types.html}
On the other hand,
the conversion is far from straightforward:
even if we could precisely check extended type correctness properties
of the C implementations,
lifting the results of the analysis to the Java-C interface
would require more legwork,
and being able to reconcile idiomatic usage in the two languages.

\paragraph{Dataflow}
As clear from \autoref{fig:ex:native}'s example,
the control flow between Java and C (or any other language compiled to native)
is bidirectional,
in that the C code can both read (e.g., \Cc{toC})
and modify (e.g., \Cc{fromC}) data in the Java runtime.
Similarly to \autoref{fig:ex:api}'s example of
Java-JavaScript API communication,
analyzing a bidirectional inter-language dataflow
is complex;
the lower-level nature of C aggravates the complexity.
In particular,
there is an array of pitfalls that may affect C programs
(such as undefined behavior, memory leaks, and so on)
and thus, indirectly, also the Java code that would
otherwise not have to worry about these issues
(thanks to Java's stricter type system, memory model,
and automatic memory management).
There is evidence that
hybrid Java-native apps
do suffer from vulnerabilities that originate in such
issues of C~\cite{JuCify2022ICSE, Ryu-Semantic-Summary, declarativeMutlilingual}.

\paragraph{Effects}
Java is an object-oriented language, and hence it
uses classes as the fundamental unit of \emph{encapsulation}.
In contrast, C does not offer comparable modular features
beyond functions (and files).
In practice, though,
C programmers often still find ways of
combining data and functions that operate on it
in a coherent way---for example,
by defining a series of functions that all take
a reference to the same \Cc{struct} as input,
as if it were the target object of methods of the same class~\cite{TFNM-ECOOP13}.
It remains that, in order
to specify and reason about programs that combine the two languages,
one needs a \emph{methodology} that
properly captures each language's notion of encapsulation---and then a way of reconciling the two formal models.
C's complex memory model,
as well as its looser notion of encapsulation,
entail that applying specification methodologies
designed for object-oriented languages like Java
to C may involve more work in terms of annotations and adaptability to
analyzing idiomatic C code.
For example, the VCC verifier~\cite{VCC} features a sophisticated ownership model
to verify low-level, idiomatic C code;
the VeriFast verifier~\cite{VeriFast}
uses separation logic and abstract predicates~\cite{ParkinsonB05}
to specify both C and Java programs in a similar style---which could be a basis
to construct a multilingual methodology.

\subsection{System Level}
\label{sec:ex:system}

\begin{figure*}[!tb]
  \centering
\begin{lstlisting}[firstline=5,lastline=13,style=displayedPy,language=Python,numbers=left]
output = subprocess.run((*\label{l:sys:run}*)
    "sort -t, -k5 -rg cities.csv | awk -F, '{print $2}'",(*\label{l:sys:utils}*)
    capture_output=True,
    shell=True(*\label{l:sys:shell}*)
)
# retrieve shell output and convert it to UTF string
data = output.stdout.decode('utf-8')(*\label{l:sys:utf}*)
# print first row of `data`
print(data.splitlines()[0])(*\label{l:sys:print}*)
\end{lstlisting}
  \caption{A Python script that communicates with shell command line utilities.}
  \label{fig:ex:system}
\end{figure*}

The lowest level of multilingual communication encompasses
very many concrete combinations of languages and systems.
As a simple example of programs that rely on system-level communication
between components implemented in different language,
we consider scripts written in Python
that interact with Unix command-line utilities
through the operating-system shell.

\autoref{fig:ex:system} shows a simple Python program
that uses standard library \Pc{subprocess}
to call the shell (line~\ref{l:sys:shell})
in order to run command-line utilities \Pc{sort} and \Pc{awk}
in combination.
Line~\ref{l:sys:utils}
displays the executed commands:
first, \Pc{sort -t, -k5 -rg cities.csv}
outputs the rows of comma-separated file \Pc{cities.csv} (\Pc{-t,})
sorted in decreasing numerical order (\Pc{-rg})
of their fifth field (\Pc{-k5});
this command's output is piped into the command
\Pc{awk -F, \'\{print $2\}\'}, which prints out the second field (\Pc{$2}) of every row.
File \Pc{cities.csv} lists information about various cities in the world;
its second column is the city name,
and its fifth column is the city's population.
Thus, the command 
returns a sequence of rows with all city names sorted from most to least populated.
Therefore,
\Pc{subprocess.run}'s call on line~\ref{l:sys:run} returns
a Python object with information about the outcome of executing the command,
such as its exit status and any error that may have occurred.
Then, line~\ref{l:sys:utf}
extracts the command's raw output (attribute \Pc{stdout})
and converts it from a byte string to a character string;
and line~\ref{l:sys:print}
splits the string at line-breaks and prints the first one
(i.e., the name of the city with the largest population count).

Even from this simple example, it is clear how the interface
between, in this case, Python and the system primitives
is low-level, and relies on several conventions to share information implicitly.

\paragraph{Types}
The range of types that are available between languages at the system level
is usually quite limited.
In our example, all the information exchanged between shell and Python program
is represented as a string
(plus possibly the numeric exit code of the shell process).
This means that there are very few checkable guarantees
that the data received by the Python program
is in a consistent format:
for example, line~\ref{l:sys:print} relies on the assumption
that \Pc{data} was properly encoded as a linebreak-separated lines,
without spurious information before the first line of data.
Conversely,
the Python program risks passing data to the shell
that is in an incorrect or inconsistent format;
for example, it has to carefully quote or escape characters that have a special
meaning in the shell (e.g., \Pc{$}).

\paragraph{Dataflow}
Precisely analyzing the dataflow
of a program that calls out to the operating system
is a daunting task.
Dynamic analysis, such as a taint analysis~\cite{10.1145/3357390.3361028, newsome2005dynamic},
is the most viable approach,
since it relies on concrete executions in a real-world environment---rather than on abstract models of the system, which would be difficult to define
accurately.
Also techniques that combine static and dynamic analysis,
such as dynamic-symbolic execution~\cite{Klee,Dart},
could be effective to detect information-flow
vulnerabilities~\cite{ThomeSBB17,Lancia22} in such hybrid programs---provided they can rely on constraint solvers
that support the ubiquitous string type~\cite{GaneshKAGHE11,Amadini23}.

\paragraph{Effects}
Reasoning about the behavior of a high-level program
that interacts directly with other system components would require a large amount of manual, expert work,
pushing the envelope of formal methods.
One approach could be to carefully define stubs
of the specific system functionalities that interact with the high-level program
(in \autoref{fig:ex:system}'s example,
the command-line utilities \Pc{sort} and \Pc{awk}),
and specify them formally;
this approach would essentially reduce the analysis problem to
a scenario of API communication (as in \autoref{sec:ex:API})
with full interface specifications.
Alternatively, and more ambitiously,
one could first re-implement the required system-level functionality
using a verified-by-construction approach such as the CakeML system~\cite{CakeML,CakeML-backend};
in this case, the binary used
to execute the system-level part of the multilingual program
would behave exactly as in its specification.

\section{Multilingual Analysis: State of the Art}
\label{sec:soa}

In tandem with the growing uptake of multilingual programming frameworks,
there is a developing interest in understanding the
limitations of ``traditional'' program analysis techniques
on multilingual programs,
and whether these programs are more likely harbor certain kinds of defects~\cite{multilanguage-smells,multilanguage-faults,multilanguage-patterns}.
In this section, we discuss the most significant work in these areas.

A language-\emph{agnostic} multilingual analysis is one
that is not overly specific to a certain combination of host and guest languages,
but is applicable, in principle, to a selection of multilingual scenarios.
In contrast, a language-\emph{specific} multilingual analysis
takes into account the specific interaction mechanism of two programming languages
within a certain programming framework---for instance, Java and JavaScript in hybrid Android apps.
Accordingly,
\autoref{sec:agnostic} discusses language-agnostic analyses,
whereas \autoref{sec:specific} discusses language-specific analyses.
Unsurprisingly, the classification in agnostic and specific
is somewhat fuzzy, in that even a language-agnostic analysis
is typically applicable only to certain kinds of multilingual frameworks
(and is demonstrated on specific instances);
however, language-agnostic analyses tend to be more abstract
than language-specific ones.

For convenience, in this section
we loosen up the host/guest terminology we have used,
in a more restricted way, elsewhere in the paper.
Namely, consider a program
where a module $H$, written in some language $\ell_H$,
calls into another module $G$, written in some other language $\ell_G$.
Then, we call $H$ the host (program); $\ell_H$ the host language;
$G$ the guest (program); and $\ell_G$ the guest language.

\subsection{Language-Agnostic Multilingual Analysis}
\label{sec:agnostic}

Lee et al.\cite{Ryu-Semantic-Summary}'s
approach works by
first summarizing the behavior of interface operations
implemented in the guest language in a way that only retains
their effects within the host language modules.
Then, any calls to these operations in the host
are replaced by their summaries,
and the host program plus summaries undergoes a standard whole program analysis.
This approach's main intuition
is that summarizing the effects at the interface with the host
is sufficient to capture the key interaction between host and guest.
Unfortunately, even such restricted summaries are challenging to construct for
realistic, feature-laden programming languages.
Besides, the precision of the whole program analysis performed
on the host program with summaries
may be reduced if the summaries fail to capture the nuances of
guest-specific program optimizations.

Prakash et al.~\cite{prakash2023unifying}
propose a language-agnostic way of combining program analyses
performed on two different languages.
Their approach first analyzes
the host and guest programs are analyzed in isolation
(ignoring inter-language communication).
Then, it builds an inter-language call graph that captures the points-to relation
across language boundaries;
this information is finally used to merge the host and guest analysis results.

Youn et al.~\cite{declarativeMutlilingual} discuss
how to apply to multilingual programs
the declarative static analyses offered by the
CodeQL vulnerability analyzer.
The basic idea is to add constraints to a declarative analysis
that capture inter-language communication.
In particular, they show
how to add constraints that approximate
the inter-language points-to relation,
so that the analysis can propagate the language-specific
constraints from host to guest and vice versa.
The main limitation of this approach is
that it is only applicable to scenarios
where exactly the same kind of analysis is available for both host and guest
and can be implemented in CodeQL.
This also excludes cases where the host and guest analysis
have different granularities
(e.g. object-sensitive vs.\ field sensitive).

Most language-agnostic frameworks
are focused on \emph{static} analysis,
which entails that they can usually only analyze a restricted set
of dynamic inter-language features.
A recent study~\cite{10.1145/3597926.3598112} suggests that
dynamic features are widely used in Android hybrid apps:
in particular, the JavaScript code run within
a WebView Android component
(the same mechanism described in \autoref{sec:ex:API})
is often passed dynamically as a string.
Another scenario that falls short of the capabilities
of language-agnostic static analysis
are inter-language communication
via Android's
intents.\URLref{https://github.com/arguslab/NativeFlowBench/tree/master/icc_javatonative}\URLref{https://github.com/arguslab/NativeFlowBench/tree/master/icc_nativetojava}
In these cases, language specific analyses have to be custom designed
to achieve a better analysis precision.

\subsection{Language-Specific Multilingual Analysis}
\label{sec:specific}

Android hybrid apps have received a lot of attention
as one of the most significant, and widespread,
scenarios of multilingual programming.
In \autoref{sec:ex:API} and \autoref{sec:ex:native},
we already outlined two pairs of languages that are
frequently used in the design of hybrid apps:
Java-JavaScript and Java-C/native.
Most program analysis work involving such apps
focuses on three major areas:
\begin{enumerate*}
\item vulnerability detection;
\item checking of information flow security properties
  (e.g. noninterference);
\item detection of type-mismatch/crashing bugs.
\end{enumerate*}
In the rest of this section, we discuss some relevant work in these areas.

\paragraph{Java-JavaScript hybrid apps}
Detecting vulnerabilities that originate
in hybrid Android apps using the WebView framework
has been the object of plenty of recent work~\cite{jsinjection, bifocal, mutchler2015large,  rizzo2017babelview, 217531, hidhaya2015supplementary, mandal2018vulnerability, Fratantonio2016LogicBomb, yang2018study, cordova, cordova2016esoss, 10172950}.

Luo et al.'s work~\cite{jsinjection}
was among the earliest demonstrations
that the WebView component can be compromised, on the Java side,
by injecting malicious JavaScript code.
Zhang et al.'s large-scale study~\cite{217531}
of web-resource manipulation APIs in hybrid apps
pointed to other kinds of vulnerabilities,
which originate in the cross-principle manipulation of web resources.
A key takeaway of this study is that
attacks may occur when an app's code manipulates web resources,
and app and resources do not originate from the same organization.
Hidhaya et al.~\cite{hidhaya2015supplementary} demonstrated a so-called
supplementary event-listener injection attack,
where an attacker registers additional event listeners---whose callbacks execute malicious activities---with the HTML elements in a webpage that is then loaded by the WebView via JavaScript injection. 

Bifocals~\cite{bifocal} is a static analysis tool
that can detect and mitigate some WebView vulnerabilities
triggered by using JavaScript interfaces that load third-party web pages.
The work on Bifocals also demonstrated man-in-the-middle attacks
that can expose sensitive data 
if we allow executing JavaScript from untrusted source.
In order to better evaluate the impact of code-injection attacks
that target WebView components,
BabelView~\cite{rizzo2017babelview}
instruments hybrid apps with an attacker model
that over-approximates the reach of possible attacks on the app's information flow.
Mandal et al.'s static analysis approach~\cite{mandal2018vulnerability}
is specialized to detect vulnerabilities in Android infotainment apps.

Misusing WebView components may also result in information-flow security bugs.
HybriDroid~\cite{hybridDroid} is a framework for detecting such type bugs;
it works by performing a taint analysis on an hybrid app's interprocedural
control-flow graph that captures the inter-language information flow.
This, in turn, is based on a formalization of
hybrid communication in Android apps.
Following a somewhat similar approach,
Bae et al.~\cite{hydridDroidICSE2019}
designed a type system that can detect bugs
originating in the misuse of Android hybrid app communication APIs.
In order to perform a large-scale study of how Android hybrid apps are programmed,
the work on LuDroid~\cite{LuDroid-Conference}
applied standard static analysis techniques
to detect all flows of information from Android into JavaScript.
They also showed that such an analysis is incomplete,
as it can miss information-flow security bugs that originate inside JavaScript;
and found numerous cases of apps that are susceptible to man-in-the-middle
attacks since they load JavaScript code dynamically.
In order to extend the range of issues that can be detected,
the same authors proposed IWandroid, a technique that
tracks sensitive data flows from Java to JavaScript and then to a public sink.
To this end, IWandroid implements an IFDE/IDE analysis~\cite{IFDE-IDE}
to summarize Java-JavaScript interface communication.
IWandroid can also detect integrity violations where the JavaScript environment modifies Android properties
(a form of bidirectional inter-language communication,
also discussed in \autoref{sec:ex:API}).

\paragraph{Java-C/native hybrid apps}
Analyzing hybrid Android apps where UI components are implemented
in C (or other language that compiles to native) and
connected to the Java host through JNI has also been a
popular research topic~\cite{ALAM2017230, 10.1145/2627393.2627396, 10.1145/2976749.2978343, Afonso2016GoingNU, 10.1145/3395363.3397368, Ryu-Semantic-Summary, 6903578, weiJNSAF-CCS2018, JuCify2022ICSE}.

JN-SAF~\cite{weiJNSAF-CCS2018}
is an efficient analysis framework for JNI hybrid Android apps,
implementing a summary-based bottom-up dataflow analysis
that captures control and dataflow behavior of native components.
More precisely, JN-SAF separately analyzes Java bytecode and binary code by symbolic analysis; then it combines the summaries of each analysis
(which use a unified abstract heap model) to detect malicious behavior
that may involve inter-language communication.
One limitation of this approach is that its heap model
is specific to the dataflow analysis JN-SAF supports;
hence it may not support other, more general analyses of program behavior.
To overcome this limitation,
JuCify~\cite{JuCify2022ICSE}
performs an analysis in multiple steps:
first, it uses symbolic execution to find calls that cross the bytecode-native
boundary;
then, it merges the call graph information about the native components
into the Java bytecode;
finally, it summarizes the semantics of the called native functions as
bytecode statements.
Underlying JuCify is a unified model of Android hybrid app code execution,
which is considerably more detailed than the previous approaches'.
Unfortunately, its detail is also a source of complexity
that limits its scalability to realistic-size apps.

\vspace{5mm}

\paragraph{Other multilingual combinations}
A few other program analysis work
targets sundry combinations of languages used together, including Python-C, Java-Python, and Lua-C~\cite{10.1007/978-3-030-88806-0_16, 279967, declarativeMutlilingual, turcotte_et_al:LIPIcs.ECOOP.2019.16, prakash2023unifying}.

Monat et al.~\cite{10.1007/978-3-030-88806-0_16} proposed an abstract-interpretation based static analysis of Python programs with C extensions. Their work leverages the Mopsa static analysis framework to combine the Python and C abstract domains; concretely,
they combined off-the-shelf value analyses for C and Python programs
by means of a custom domain that gives an abstract semantic to operations
at the inter-language boundary.
PolyCruise~\cite{279967} is another framework that can analyze Python-C programs. It is based on a dynamic flow analysis to find inter-language dependencies
sources and sinks of sensitive data.
PolyCruise can also be used \emph{online}, while a program is running,
to monitor its behavior and find vulnerabilities
that may only occur in operating conditions.

Turcotte et al.~\cite{turcotte_et_al:LIPIcs.ECOOP.2019.16} proposed a framework for analyzing Lua-C programs based on a simplified model of the guest language (i.e., C).
To this end,
it combines the host and guest type systems at the inter-language interface.
While this type model abstracts away several details of the inter-language semantics,
it fully captures statically any possible type errors that may ensue.
Furthermore, after type checking determined that the inter-language
interface is type safe, one can remove dynamic wrappers in Lua code,
which contributes to improving the performance of the program.
The approach is fairly general; however, applying it
to realistic languages still requires some extra work,
as one has to define a fully-typed foreign-function interface between languages
(for example, the paper uses Poseidon Lua, a variant of Lua that offers gradual typing).

\section{Discussion}
\label{sec:discussion}

In the paper, we discussed several challenges of multilingual program
specification and analysis,
and summarized some of the research in this area in the last decade or so.

A lot of progress has occurred \emph{bottom-up}:
the starting point were empirical studies
(or even ad hoc observations)
of concrete vulnerabilities that can be found in
certain multilingual programs.
The realization that a particular misbehavior may occur
at the interface between two language runtimes
motivated researchers to come up with bespoke combinations
of existing program analyses that capture the semantics
of such inter-language communication,
and hence can detect the vulnerabilities automatically
and on a large scale.
Even though it is somewhat piecemeal,
the bottom-up approach will remain fruitful in the future,
as it caters to concrete, specific challenges---which are therefore more likely to be manageable,
and have a clear practical relevance.

On the other hand,
there is also room for more research in multilingual program analysis
that is \emph{top-down}:
starting from the state of the art in modeling and reasoning about
the semantics of \emph{one} standalone programming language,
researchers can explore extensions and combinations that
cover the additional ground involving inter-language communication.
An interesting direction for future work following
the top-down approach could be leveraging
frameworks that have been used to rigorously model
the semantics of programming languages---one at a time---in particular ``lightweight'' techniques
such as the K framework~\cite{K-fram}\URLref[The K Semantic Framework]{https://kframework.org/} and
PLT Redex~\cite{PLT-Redex}\URLref[PLT Redex]{https://redex.racket-lang.org/}.
A top-down approach could push the boundaries of multilingual program analysis
in a way that goes beyond the immediate, practical issues
that are important for practitioners.


\printendnotes[itemize]

\end{document}